\documentclass[aps,prd,12pt,nofootinbib]{revtex4-2}
\usepackage{slashed}
\usepackage{dutchcal}
\usepackage{graphicx}
\usepackage{amsmath}
\usepackage{amssymb}
\usepackage{mathrsfs}
\usepackage{verbatim}
\usepackage{comment}

\usepackage[normalem]{ulem}
\usepackage{xcolor}

\newcounter{fig}   \newcommand{\lbfig}[1]{\refstepcounter{fig}
\label{#1} }

\newcommand{\Tr}{{\rm Tr}}

\newcommand{\bea}{\begin{eqnarray}}
\newcommand{\eea}{\end{eqnarray}}
\newcommand{\be}{\begin{equation}}
\newcommand{\ee}{\end{equation}}

\def\({\left(}
\def\){\right)}

\newcommand{\re}[1]{(\ref{#1})}

\def\rlx{\relax\leavevmode}
\def\IR{\rlx\hbox{\rm I\kern-.18em R}}
\def\one{\hbox{{1}\kern-.25em\hbox{l}}}

\newcommand{\eqn}{\begin{eqnarray}}
\newcommand{\eqnx}{\end{eqnarray}}

\date{\today}

\begin{document}
\title{
Fermion states localized \\ on a self-gravitating non-Abelian monopole}
\author{
Vladimir Dzhunushaliev
}
\email{v.dzhunushaliev@gmail.com}
\affiliation{
Department of Theoretical and Nuclear Physics,  Al-Farabi Kazakh National University, Almaty 050040, Kazakhstan
}
\affiliation{
Institute of Experimental and Theoretical Physics,  Al-Farabi Kazakh National University, Almaty 050040, Kazakhstan
}
\affiliation{
Academician J.~Jeenbaev Institute of Physics of the NAS of the Kyrgyz Republic, 265 a, Chui Street, Bishkek 720071, Kyrgyzstan
}

\author{Vladimir Folomeev}
\email{vfolomeev@mail.ru}
\affiliation{
Institute of Experimental and Theoretical Physics,  Al-Farabi Kazakh National University, Almaty 050040, Kazakhstan
}
\affiliation{
Academician J.~Jeenbaev Institute of Physics of the NAS of the Kyrgyz Republic, 265 a, Chui Street, Bishkek 720071, Kyrgyzstan
}

\author{Yakov Shnir}
\email{shnir@theor.jinr.ru}
\affiliation{BLTP, JINR, Dubna 141980, Moscow Region, Russia}
\affiliation{
Institute of Physics,
Carl von Ossietzky University of Oldenburg,
Oldenburg D-26111, Germany}

\begin{abstract}
We study fermionic modes localized on the static spherically symmetric  self-gravitating non-Abelian monopole in the
$SU(2)$ Einstein-Dirac-Yang-Mills-Higgs
theory. We consider dependence of the spectral flow on the effective gravitational coupling constant and  show that,
in the limiting case of transition to the Reissner-Nordstr\"{o}m black hole, the fermion modes are fully absorbed into
the interior of the black hole.
\end{abstract}
\maketitle

\section{Introduction}
Various black holes  with localized matter fields,
which circumvent the no-hair theorem
(see, e.g., \cite{Volkov:1998cc,Herdeiro:2015waa,Volkov:2016ehx} and references therein),
are rather a common presence in the landscape of gravity solutions. The most well-known examples
in (3+1)-dimensional asymptotically flat spacetime
are static hairy black holes with spherically symmetric event horizon
in the $SU(2)$ Einstein-Yang-Mills  theory \cite{Volkov:1989fi,Volkov:1990sva,Bizon:1990sr},
black holes with Skyrmion hairs \cite{Luckock:1986tr,Droz:1991cx,Bizon:1992gb} and black holes inside magnetic monopoles
\cite{Lee:1991vy,Breitenlohner:1991aa,Breitenlohner:1994di}. Various generalizations of solutions of that type with different types of hairs
were considered over last decade. In particular, there are spinning black holes with
scalar hairs both in the Einstein-Klein-Gordon theory \cite{Hod:2012px,Herdeiro:2014goa} and in the non-linear O(3) sigma model
\cite{Herdeiro:2018djx}, dyonic black holes in Einstein-Yang-Mills-Higgs theory \cite{Brihaye:1999nn,Brihaye:1998cm} and black holes with axionic
hairs \cite{Campbell:1990ai,Delgado:2020hwr}. There are also hairy black holes supporting the stationary Proca hair
\cite{Herdeiro:2016tmi} and electrostatic charged black holes
\cite{Herdeiro:2020xmb,Hong:2019mcj,Kunz:2023qfg}.

In most cases, such solutions can be viewed as a small black hole immersed inside a localized field configuration, the horizon radius $r_h$
cannot be arbitrary large. The limiting case of the event horizon shrinking to zero corresponds to the regular self-gravitating lump,
which may also possess a flat space solitonic limit. The corresponding solutions may represent a topological soliton, like monopoles
\cite{tHooft:1974kcl,Polyakov:1974ek} and Skyrmions \cite{skyrme,Skyrme:1962vh}, or a non-topological solitons, like
Q-balls \cite{Rosen:1968mfz,Friedberg:1976me,Coleman:1985ki}. There is also another class of spinning hairy black holes which do not possess the
solitonic limit, like black holes with stationary Klein-Gordon hair \cite{Hod:2012px,Herdeiro:2014goa} or black holes with Yang-Mills hair
\cite{Volkov:1989fi,Volkov:1990sva,Bizon:1990sr}.

On the other hand, some of hairy black holes with finite horizon radius
may bifurcate with the vacuum black holes, as it happens, for example, with the black holes with monopole hair, they smoothly approach
the extremal Reissner-Nordstr\"{o}m solution~\cite{Lee:1991vy,Breitenlohner:1991aa,Breitenlohner:1994di,Lue:1999zp}. Another scenario is that there is a mass gap between
hairy black holes and corresponding vacuum solutions with an event horizon. This situation  takes
place for black holes with Skyrmion hairs
\cite{Luckock:1986tr,Droz:1991cx,Bizon:1992gb} and for the black holes with pure Yang-Mills hairs on
the Schwarzschild background
\cite{Volkov:1989fi,Volkov:1990sva,Bizon:1990sr}.

A notable exception in the variety of asymptotically flat solutions of General Relativity in (3+1) dimensions,
which circumvents the no-hair theorems,
is a missing class of black holes with {\it{fermionic}} hairs. Although there are regular localized solutions of the Einstein-Dirac and
Einstein-Maxwell-Dirac equations
\cite{Finster:1998ws,Finster:1998ux,Herdeiro:2019mbz,Herdeiro:2017fhv,Dzhunushaliev:2018jhj,Dzhunushaliev:2019kiy,Herdeiro:2021jgc}, all  attempts to
extend these solutions to the case of
finite event horizon has been failed: the spinor modes, which are gravitationally bound in the black hole spacetime
decay due to the absence of superradiance mechanism for the Dirac field \cite{Dolan:2015eua}.
On the other hand, black holes with fermionic hairs are known to exist in the gauged $d=4,5$ $N=2$ supergravity
\cite{Burrington:2004hf,Hristov:2010eu}; here the $N=2$ extremal black holes represent $1/2$ Bogomolnyi-Prasad-Sommerfield (BPS)
states \cite{Ferrara:1995ih} with a set of fermion zero modes. Appearance of these modes is related with remarkable relation between the
topological charge of the field configuration and the number of zero modes, exponentially localized on a soliton:
the fundamental Atiyah-Patodi-Singer index theorem \cite{Atiyah:1975jf}
requires one normalizable fermion zero mode per unit topological charge. Recently, massless electroweak fermions in the near horizon
region of black hole we discussed in Ref.~\cite{Maldacena:2020skw}.

The fermion modes localized on a soliton are well known and are exemplified by
the spinor modes of
the kinks \cite{Dashen:1974cj,Jackiw:1975fn}, vortices \cite{Fermivortex,Jackiw:1981ee},
Skyrmions \cite{Hiller:1986ry,Kahana:1984dx}
and monopoles \cite{Jackiw:1975fn,Rubakov:1982fp,Callan:1982au}. In supersymmetric theories the fermion zero
modes are generated via supersymmetry transformations of the boson field of the static soliton; breaking of
supersymmetry yields a spectral flow of the eigenvalues
of the Dirac operator with some number of normalizable bounded modes crossing zero.
However, little is known about evolution of the bounded
fermionic modes in the presence of gravity, especially as the self-gravitating soliton
approaches the critical limit and bifurcates with a
black hole.

In this Letter we investigate numerically a self-gravitating non-Abelian monopole-fermion system with back-reaction and elucidate the
mechanism for disappearance of the fermionic modes. Our computations reveal that as the BPS monopole bifurcates with the
extremal Reissner-Nordstr\"{o}m solution, the fermionic modes become absorbed into the interior of the black hole. Further, we show that
this observation also holds for non-BPS monopoles with localized non-zero modes.

\section{The Model}
We consider the (3+1)-dimensional $SU(2)$ Einstein-Yang-Mills-Higgs system, coupled to a
spin-isospin field $\psi_{\alpha i}$. The model has the following action
(we use natural units with $c=\hbar=1$ throughout):
\begin{equation}
S=\int d^4x~\sqrt{-g}\left[
-\frac{R}{16 \pi G}  - \frac{1}{2}\Tr (F_{\mu \nu} F^{\mu \nu})
    + \Tr (D_\mu \phi~ D^\mu \phi)
    - \frac{\lambda}{4} \Tr \left(
        \phi^2 - \phi_0^2
    \right)^2
    + L_{\text{sp}}\, \right],
\label{lgr_10}
\end{equation}
where $R$ is the scalar curvature, $G$ is Newton's gravitational constant, $g$
denotes the determinant of the metric tensor,  and the field strength tensor
of the  gauge field $A_\mu=\frac12 A_\mu^a \tau^a $ is
$$
F_{\mu\nu}=\partial_\mu A_\nu - \partial_\nu A_\mu + ie [A_\mu, A_\nu] \, ,
$$
where $a=1, 2, 3$ is a color index, $\mu, \nu = 0, 1, 2, 3$ are spacetime indices, and
$\tau^a$ are the Pauli matrices.
The covariant derivative of the scalar field in adjoint representation
$\phi=\phi^a \tau^a$ is
$$
D_\mu \phi = \partial_\mu\phi + ie [A_\mu,\phi] ,
$$
where $e$ is the gauge coupling constant.
The scalar potential with a Higgs vacuum expectation value  $\phi_0$
breaks the $SU(2)$  symmetry down to $U(1)$ and the scalar self-interaction constant $\lambda$ defines the mass of the
Higgs field, $M_s=\sqrt \lambda \phi_0$. The gauge field becomes
massive due to the coupling with the scalar field, $M_v=e\phi_0$.

Bosonic sector of the model $\re{lgr_10}$ is coupled to the Dirac
isospinor fermions $\psi_{\alpha i}$ with the Lagrangian \cite{Jackiw:1975fn}
\be
L_{\text{sp}}=  \frac{\imath}{2}\left(
(\hat {\slashed{D}}\bar \psi) \psi - \bar \psi \hat {\slashed D} \psi \right) - m\bar \psi \psi -
\frac{\imath }{2}h \bar \psi \gamma^5 \phi \psi \, ,
\label{Lag-spinor}
\ee
where $m$ is a bare mass of the fermions, $h$ is the Yukawa coupling constant,
$\gamma^\mu$ are the Dirac matrices in the standard representation in a curved spacetime,
$\gamma^5$ is the corresponding Dirac matrix defined in Appendix~\ref{append1},
$\hat {\slashed D} = \gamma^\mu \hat D_\mu $  and
the isospinor covariant derivative on a curved spacetime is defined as (see, e.g., Ref.~\cite{Dolan:2015eua})
$$
\hat D_\mu \psi = (\partial_\mu - \Gamma_\mu + ie A_\mu)\psi .
$$
Here $\Gamma_\mu$ are the spin connection matrices~\cite{Dolan:2015eua}.
Explicitly, in component notations, we can  write
$$
\hat D_\mu \psi_{\alpha i}  \equiv \left[
\delta_{ij}(\partial_{\mu} - \Gamma_\mu )
- \frac{\imath e}{2}(\tau^a)_{ij}
        A^a_\mu\right]\psi_{\alpha i}
$$
with the group
indices $i,j$ taking the values
$1, 2$ and the Lorentz index $\alpha$ takes the values $0\dots 3$.

Variation of the action  \re{lgr_10} with respect to the metric leads to the Einstein  equations
\be
R_{\mu \nu} - \frac{1}{2} g_{\mu \nu}R = 8\pi G
\left[ \left( T_{\mu \nu} \right)_{YM} + \left( T_{\mu \nu} \right)_{\phi} + \left( T_{\mu \nu} \right)_{s}
\right]
\label{Einst-eq}
\ee
with the pieces of the total stress-energy tensor
\be
\begin{split}
\left( T_{\mu \nu} \right)_{YM} = & - F^a_{\mu \alpha} F^a_{\nu \beta}g^{\alpha \beta}
    + \frac{1}{4} g_{\mu \nu}  F^2\, ,\\
\left( T_{\mu \nu} \right)_{\phi} = & D_\mu \phi^a D_\nu \phi^a
    - g_{\mu \nu} \left[ \frac{1}{2}D_\alpha \phi^a D^\alpha \phi^a - \frac{\lambda}{4} \left(
        \phi^2  - \phi_0^2
    \right)^2\right]\, , \nonumber\\
\left( T_{\mu \nu} \right)_{s}  =&\frac{\imath}{4}
\left[\bar\psi \gamma_{\mu} (\hat D_\nu \psi) +\bar\psi\gamma_\nu (\hat D_\mu\psi)
-(\hat D_\mu \bar  \psi) \gamma_{\nu }\psi- (\hat D_\nu \bar\psi) \gamma_\mu\psi
\right]-g_{\mu\nu} L_{\text{sp}}\, .
\end{split}
\ee
The corresponding matter field equations are:
\be
\begin{split}
D_\nu F^{a \nu \mu}& =  -e \epsilon^{abc} \phi^b D^\mu \phi^c
    - \frac{e}{2} \bar \psi \gamma^\mu \sigma^a \psi \, , \\
D_\mu D^\mu \phi^a &+ \lambda \phi^a \left(\phi^2  - \phi_0^2 \right)
    + \imath h \bar \psi \gamma^5 \sigma^a \psi =  0 \, ,\\
\imath \hat {\slashed{D}} \psi & -\imath \frac{ h}{2} \gamma^5 \sigma^a \phi^a \psi - m \psi =  0 .
\label{field_eqs}
\end{split}
\ee

\section{Equations and solutions}

Working within the above model, in this section we present general spherically symmetric equations and solve them numerically
for some values of system parameters.

\subsection{The Ansatz}
For the gauge and Higgs field we employ the usual static spherically symmetric
hedgehog Ansatz \cite{tHooft:1974kcl,Polyakov:1974ek}
\be
A_0^a=0,\quad A_i^a= \varepsilon_{aik} \frac{r^k}{er^2} \left[ 1 - W(r) \right],\qquad
\phi^a=\frac{r^a}{e r}H(r)\,.
\label{fields-boson}
\ee

The spherically symmetric Ansatz with a harmonic time dependence
for the
isospinor fermion field localized by the monopole can be written
in terms of  two $2\times 2$ matrices $\chi$ and $\eta$~\cite{Jackiw:1975fn,Jackiw:1976xx} as
$$
\psi =  e^{-\imath\omega t}\begin{pmatrix}
        \chi \\
        \eta
    \end{pmatrix}
\quad \text{with} \quad
 \chi =
    \frac{u(r)}{\sqrt{2}} \begin{pmatrix}
        0   &   -1 \\
        1   &   0
    \end{pmatrix},
  \eta =
     \imath \frac{v(r)}{\sqrt{2}} \begin{pmatrix}
        \sin \theta e^{- \imath \varphi}    &   - \cos \theta \\
        - \cos \theta    &   -\sin \theta e^{\imath \varphi}
    \end{pmatrix} .
$$
Here $u(r)$ and $v(r)$ are two real functions of the radial coordinate only and $\omega$ is the eigenvalue of the Dirac operator.

For the line element we employ Schwarzschild-like coordinates, following closely the usual
consideration of gravitating monopole (see, e.g., Refs.~\cite{Breitenlohner:1991aa,Breitenlohner:1994di})
\begin{equation}
    ds^2 = \sigma(r)^2 N(r) dt^2 - \frac{dr^2}{N(r)} - r^2 (d\theta^2 + \sin^2 \theta d\varphi^2) .
\label{metrics}
\end{equation}
The metric function $N(r)$ can be rewritten as $N(r)=1-\frac{2 G \mu(r)}{r}$ with the mass function $\mu(r)$;  the ADM mass of the
configuration is defined as $M=\mu(\infty)$.
The above metric implies the following form of the orthonormal tetrad:
$$
    e^a_{\phantom{a} \mu} = \text{diag} \left\lbrace
        \sigma \sqrt{N}, \frac{1}{\sqrt{N}}, r, r \sin \theta
    \right\rbrace ,
$$
such that $ds^2=\eta_{ab}(e^a_\mu dx^\mu )(e^b_\nu dx^\nu)$, where the Minkowski metric $\eta_{ab}=(+1,-1,-1,-1)$ and
$\gamma^\mu= e^\mu_{\phantom{a} a} \hat \gamma^a$ with $\hat \gamma^a$ being the usual
flat space Dirac matrices.

\subsection{Equations}

Substitution of
the Ansatz \re{fields-boson}-\re{metrics}
into the general system of equations \re{Einst-eq} and \re{field_eqs} yields the following
set of six coupled ordinary differential equations for the functions $W,H,u,v,N,\sigma$
(here the prime denotes differentiation  with respect to the radial coordinate, $\sigma ' = \frac{d\sigma}{dr}$, etc.
):
\begin{align}
    &
    \frac{\sigma '}{\sigma } =  \alpha^2  \left[
    2\frac{W^{\prime 2}}{x } + x H^{\prime 2} - \frac{2 W + h x  H}{N }u v
    + 2\omega \frac{ x\left(u^2 + v^2\right)}{N^{3/2}\sigma}
    - m \frac{x \left(u^2-v^2\right)}{N}
    \right]\, ,
\label{field_eqs_curved_Einstein_1}\\
    &
    N' + \frac{1}{x}\left(N - 1\right) =  - \alpha^2  \Big[
    2\frac{N W^{\prime 2}}{x}
    + x N H^{\prime 2}
    + \frac{\left(1 - W^2\right)^2}{x^3}
    + 2\frac{W^2 H^2}{x}
    + \frac{\beta^2}{2} x \left(1-H^2\right)^2 \nonumber \\
    &+ 2\omega\frac{x \left(u^2 + v^2\right)}{\sqrt{N}\sigma}
    \Big]\, ,
\label{field_eqs_curved_Einstein_2}\\
    &
    W''+ \left(
        \frac{N'}{N} + \frac{\sigma'}{\sigma}
    \right)W'
    + \frac{\left(1- W^2\right) }{N x^2}W
    = \frac{W H^2}{N } + \frac{ x u v}{N}\, ,
\label{YM_eqs}\\
    &
    H'' + \left(
        \frac{2}{x} + \frac{N'}{N}
    + \frac{\sigma'}{\sigma}
    \right)H'
    - 2\frac{ W^2 H}{N x^2} + \frac{\beta^2}{N} \left(1 - H^2\right)H
    - 2 h \frac{u v}{N}= 0 \, ,
\label{phi_eqn}
\end{align}
\begin{align}
    &
    u' + u \left(
    - \frac{W}{\sqrt{N} x} - \frac{h}{2}\frac{ H}{\sqrt{N}}
    + \frac{1}{4}\frac{N'}{N}
    + \frac{1}{x} + \frac{1}{2}\frac{\sigma'}{\sigma}
    \right)
    + v \left( \frac{\omega}{N\sigma} + \frac{m}{\sqrt{N}} \right) = 0 \, ,
\label{psi_eqn_1}   \\
    &
    v' + v \left(
        \frac{W}{\sqrt{N} x} + \frac{h}{2}\frac{H}{ \sqrt{N}} + \frac{1}{4}\frac{N'}{N}
        + \frac{1}{x} + \frac{1}{2}\frac{\sigma'}{\sigma}
    \right)
    - u \left( \frac{\omega}{N\sigma} + \frac{m}{\sqrt{N}}\right) = 0\, .
\label{psi_eqn_2}
\end{align}
Here we define a new dimensionless radial coordinate, $x=e \phi_0 r$, and three rescaled  effective
coupling constants $\alpha^2=4\pi G\phi_0^2\,, \beta^2=\frac{\lambda}{e^2}\, , \tilde h=\frac{h}{e}$.
The scaled bare mass parameter and the eigenfrequency of the fermion field are
$\tilde{m}=\frac{m}{e \phi_0}$ and $\tilde{\omega}=\frac{\omega}{g\phi_0}$, respectively. The fermion field
scales as $\psi \to \psi/(\sqrt{e}\phi_0^{3/2})$. To simplify the formulas, we will  drop
the tilde notation henceforth.  Also, in what follows, we restrict ourselves to the
case of  fermions with zero bare mass setting $m=0$.
Hence, the solutions depend essentially on three dimensionless parameters given by the mass ratios
$$
\alpha=\sqrt{4\pi}\frac{M_v}{e M_{Pl}}, \quad \beta = \frac{M_s}{M_v}, \quad h = \frac{2M_f}{M_v} ,
$$
where $M_{Pl}= G^{-1/2}$ is the Plank mass and
$M_v=e\phi_0$,  $M_s=\sqrt{\lambda} \phi_0$ and $M_f=h \phi_0/2$ are the masses of the
gauge field, Higgs field and fermion field, respectively.

The system of equations \re{field_eqs_curved_Einstein_1}-\re{psi_eqn_2} is supplemented
by the normalization condition of the localized fermion mode\footnote{In our numerical calculations we fix $e=0.689$.
}
\be
  \int dV\, \psi^\dagger \psi =
\frac{4\pi}{e^{2}} \int_{0}^{\infty}\frac{\tilde{u}^2 + \tilde{v}^2}{\sqrt{N}}x^2 dx
=    1 .
\label{norm}
\ee

Note that, as $\omega \neq 0$, the metric field
$\sigma$ cannot be eliminated from the system~\re{field_eqs_curved_Einstein_1}-\re{psi_eqn_2},
as is done, for example, for a self-gravitating monopole (see, e.g., Ref.~\cite{Volkov:1998cc}).

The system \re{field_eqs_curved_Einstein_1}-\re{psi_eqn_2} admits embedded Reissner-Nordstr\"{o}m (RN) solution \cite{Bais:1975gu,Cho:1975uz}; for the case
of unit magnetic charge it reads
\be
\sigma=1\, , \quad \mu(x)=\mu_\infty - \frac{\alpha^2}{2x}\, ,\quad W=0\, , \quad H=1\, , u=v=0 .
\label{RN}
\ee
A horizon occurs when $N(x) \to 0$, in the Schwarzschild-like parametrization it happens at
some finite critical value of $x=x_{\text{cr}}=\alpha_{\text{cr}}$.

\subsection{Numerical results}

The system \re{lgr_10} possesses two limits. The flat space monopole corresponds to the case $\alpha=0$;
further, setting $\beta=0$, yields the familiar self-dual BPS solution~\cite{Bogomolny:1975de,Prasad:1975kr} (see also
Ref.~\cite{Shnir:2005vvi} for a review),
\be
W(x)=\frac{x}{\sinh x}\, ,\qquad H(x)=\coth x - \frac{1}{x}.
\label{BPS-boson}
\ee
There is a remarkable flat space solution for the background isospin fermion zero ($\omega=0$) mode
\cite{Jackiw:1975fn,Jackiw:1976xx}. Indeed, in this case the last pair of equations~\re{field_eqs_curved_Einstein_1}-\re{psi_eqn_2} is decoupled, and it is reduced to
\be
\begin{split}
    u' + u \left(
    \frac{1-W}{ x} - \frac{h}{2} H\right) &= 0 \, , \nonumber\\
    v' + v \left(
        \frac{1 + W}{ x} + \frac{h}{2}H\right) &= 0 .
\end{split}
\ee
Using the vacuum boundary conditions, we can see that the linearized asymptotic equations for the spinor components
approaching the vacuum are
$$
u' -\frac{hu}{2} + \omega v\approx 0\, ,\qquad
v' + \frac{hv}{2} - \omega u\approx 0 .
$$
Therefore, gravitationally localized
fermion modes with exponentially decaying tail may exist if $\omega^2 < h^2/4$.

The normalizable solution for the localized zero mode is
$$
v=0\, , \qquad u\sim \exp\left\{-\int dx^\prime\, \left[\frac{1-W(x^\prime)}{x^\prime} -
\frac{h}{2}H(x^\prime)\right] \right\} \, ,
$$
and it exists only for non-zero negative values of the scaled Yukawa coupling $h$. For example, setting
$h=-2$ and making use of the exact
BPS monopole solution \re{BPS-boson},
we obtain
\be
v=0\, , \qquad u=\frac{1}{\cosh^2(x/2)}\, .
\label{BPS-fermion}
\ee
In our numerical calculations we used these closed form BPS solutions as a input.

Another limit $h\to 0$ while $\beta$ is kept fixed, corresponds to the decoupled fermionic sector. In such a case
the  well known pattern of evolution of the self-gravitating monopole is recovered, a branch of
gravitating solutions emerges smoothly from the flat space monopole
as the effective gravitational coupling $\alpha$ increases from zero and  $\beta$ remains fixed
\cite{Lee:1991vy,Breitenlohner:1991aa,Breitenlohner:1994di}. Along this branch the metric function $N(x)$
develops a minimum, which decreases monotonically.
The branch terminated at a critical value $\alpha_{\text{cr}}$ at which the gravitating monopole develops
a degenerate horizon and configuration collapses into the extremal Reissner-Nordstr\"{o}m black hole,
as displayed in the left panel of Fig.~\ref{fig1}. A short backward branch
of unstable solutions arises in the BPS limit $\beta=0$ at $\alpha=\alpha_{\text{max}}$,
it bends backwards and bifurcates with the branch of extremal RN solutions of unit magnetic charge
at $\alpha_{\text{cr}}< \alpha_{\text{max}}$ \cite{Lee:1991vy}.
Note that the ADM mass of the monopole coupled to the fermion zero mode remains the same
as the mass of the pure self-gravitating monopole; this is because
the non-zero spinor component $u(x)$ is decoupled and there is no backreaction of the fermions, see below.

\begin{figure}[t!]
    \begin{minipage}[t]{.49\linewidth}
        \begin{center}
\includegraphics[width=.98\linewidth]{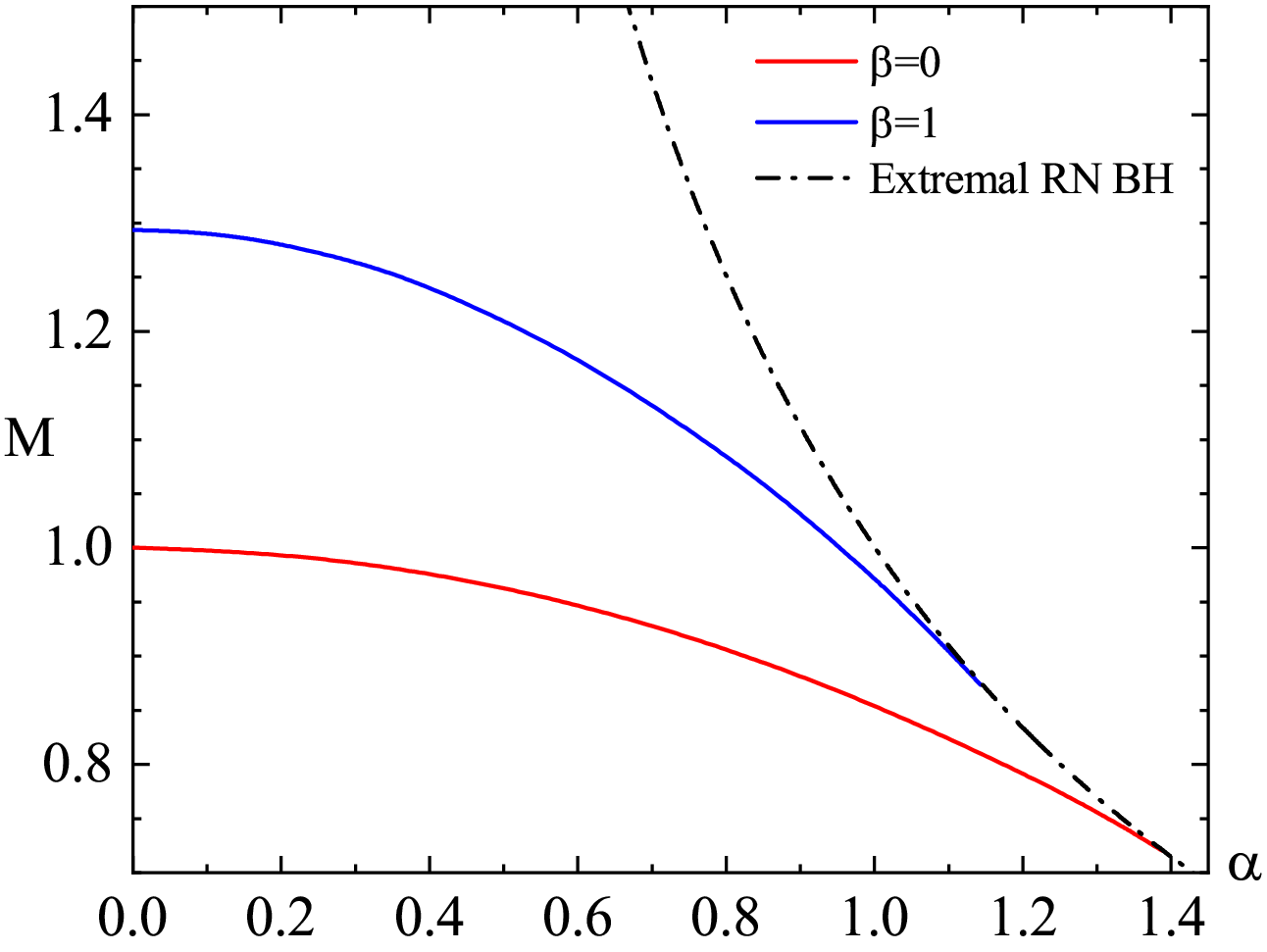}
        \end{center}
    \end{minipage}\hfill
    \begin{minipage}[t]{.49\linewidth}
        \begin{center}
\includegraphics[width=1\linewidth]{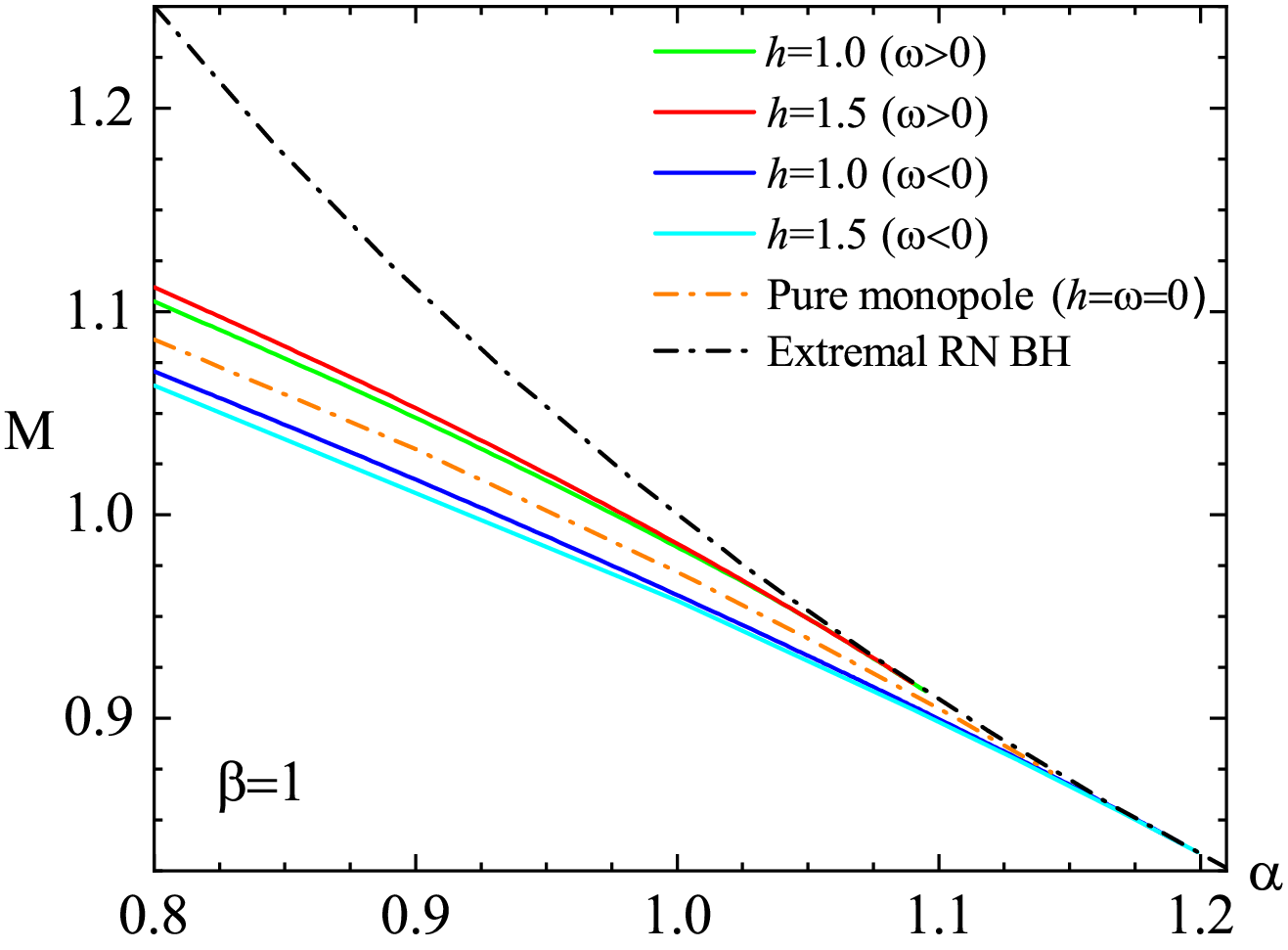}
\end{center}
    \end{minipage}
\caption{Left panel:
The dependence of the ADM mass $M$ of the gravitating monopole on the effective
gravitational coupling $\alpha$ is
shown for $\beta=0$ and $\beta=1$ at $h=-1$ and  $\omega=0$.
Right panel: The same dependence is shown for the bounded monopole-fermion
system with nonzero (positive and negative) eigenvalues $\omega$
   for $\beta=1$ and $h=1, 1.5$.
For comparison, in both panels, the mass of the extremal Reissner-Nordstr\"{o}m black hole of unit charge is also shown.
}
\lbfig{fig1}
\end{figure}

Generally, the system of mixed order differential equations \re{field_eqs_curved_Einstein_1}-\re{psi_eqn_2} can be solved numerically together with constraint
imposed by the normalization condition \re{norm}. The boundary conditions are
found by considering the asymptotic expansion of the solutions on the boundaries of the
domain of integration together with the assumption of regularity and asymptotic flatness.
Explicitly, we impose
\be
\begin{split}
N(0)=&1, \quad W(0)=1,\quad H(0)=0,\quad v(0)=0, \quad \partial_x u(0)=0,
\quad \partial_x \sigma(0)=0;\\
N(\infty)=&1, \quad W(\infty)=0,\quad H(\infty)=1,\quad v(\infty)=0, \quad u(\infty)=0,
\quad \sigma(\infty)=1 \, .
\label{BC}
\end{split}
\ee

Consider first the evolution of the fermion zero mode localized on the self-gravitating BPS monopole. Note that
since both the bare mass of the fermion field and the eigenvalue of the Dirac operator are zero,
there is no backreaction of the fermions on the monopole, the system of equations~\re{field_eqs_curved_Einstein_1}-\re{psi_eqn_2} becomes decomposed into 3 familiar
coupled equations for self-gravitating monopole \cite{Lee:1991vy,Breitenlohner:1991aa,Breitenlohner:1994di}
and two decoupled equations for the components of the localized fermion mode.
\begin{figure}[t!]
\begin{center}
\includegraphics[width=1.\linewidth]{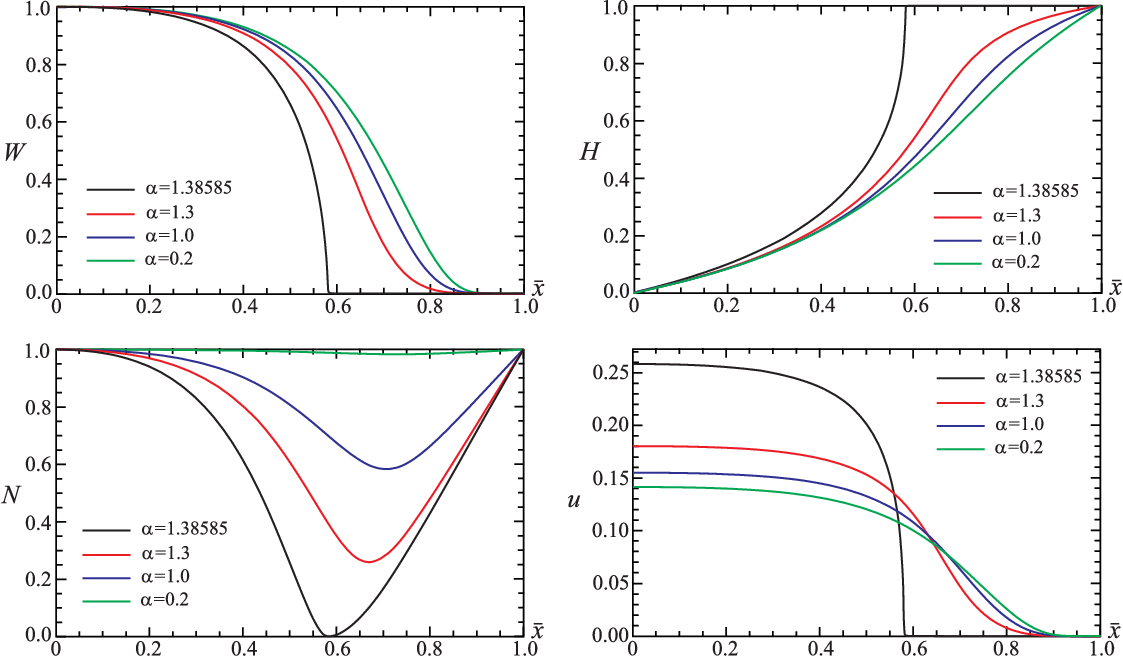}
\end{center}
\vspace{-0.5cm}
\caption{
The  profile functions of the solutions of the system~\re{field_eqs_curved_Einstein_1}-\re{psi_eqn_2} in the BPS limit $\beta=0$
are shown as functions of the compactified radial  coordinate $\bar x=x/(1+x)$ for some set of values of the
effective gravitational coupling $\alpha$ at  $\omega=0$ and $h=-1$.
The spinor component $v$ always remains zero.
}
\lbfig{fig2}
\end{figure}

The fundamental branch of gravitating BPS monopoles with bounded fermionic zero mode smoothly arise
from the flat space configuration \re{BPS-boson},\re{BPS-fermion}
as the effective gravitational constant $\alpha$ is increased above zero. This branch
reaches a limiting solution at maximal value $\alpha_{\text{max}}=1.403$, here it bifurcates with the short backward branch
which leads to the extremal RN black hole with unit magnetic charge, see Fig.~\ref{fig1}.

In Fig.~\ref{fig2} we displayed the corresponding solutions for some set of values of the
effective gravitational coupling $\alpha$ at $h=-1$ and $\beta=0$. With increasing $\alpha$
the size of the configuration with localized modes is gradually decreasing.
As the critical value of $\alpha$ is approached, the minimum of the metric function $N(x)$
tends to zero at $x=x_{\text{cr}}$. The metric becomes splitted into the inner part, $x< x_{\text{cr}}$ and
the outer part, $x> x_{\text{cr}}$ separated by the forming hozizon. The Higgs field is taking the vacuum expectation value
in exterior of the black hole while the gauge field profile function $W(x)$ trivializes there,
so the limiting configuration corresponds
to the embedded extremal RN solution \re{RN} with a
Coulomb asymptotic for the magnetic field. At the same time, the fermion field becomes absorbed into the
interior of the black hole, see Fig.~\ref{fig2}.

\begin{figure}[t!]
\begin{center}
    \includegraphics[width=1.\linewidth]{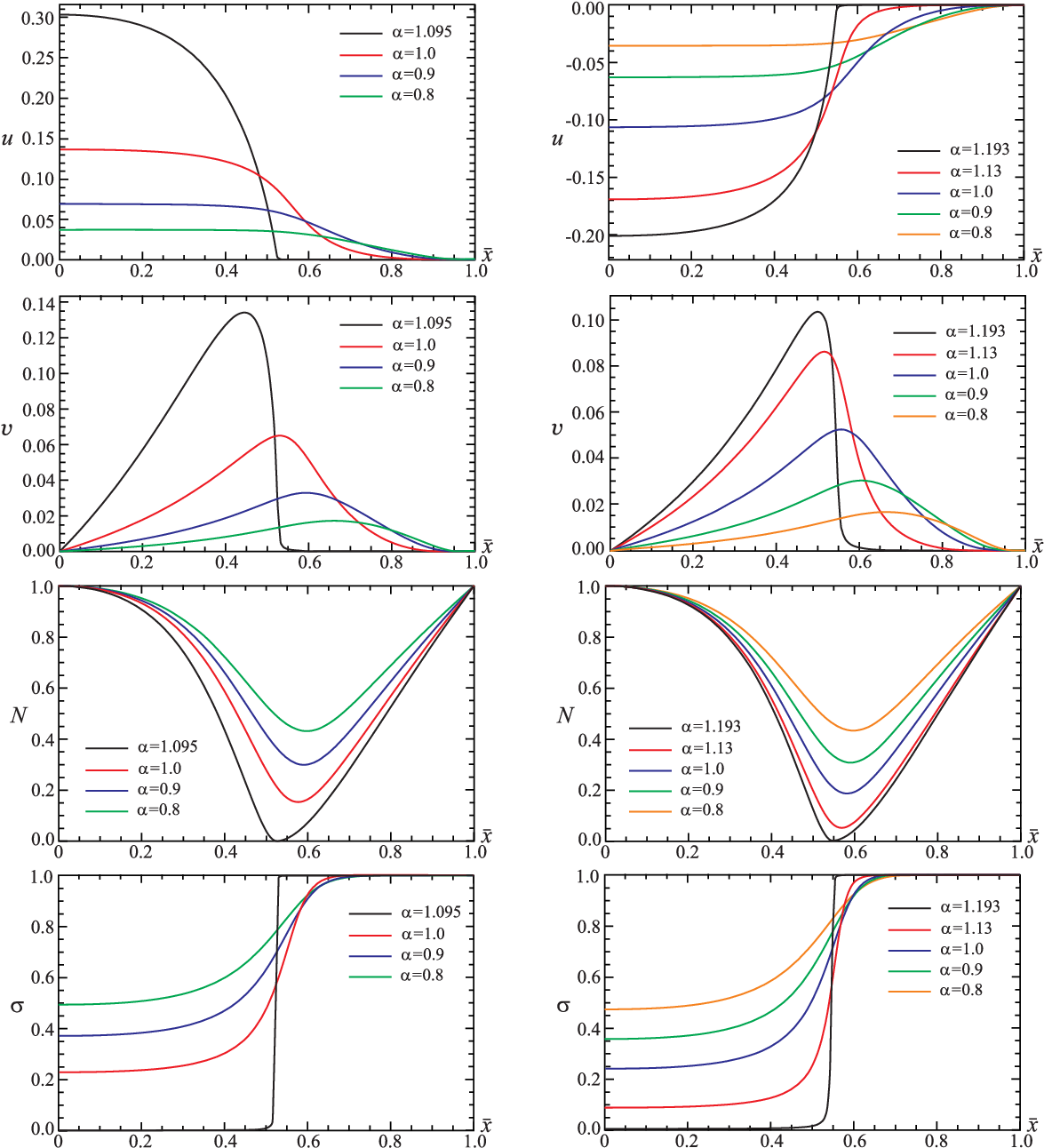}
\end{center}
\vspace{-0.5cm}
    \caption{
        The profiles of the spinor and metric functions of the solutions of the system~\re{field_eqs_curved_Einstein_1}-\re{psi_eqn_2}
are shown as functions of the compactified radial  coordinate $\bar x=x/(1+x)$ for some set of values of the
effective gravitational coupling $\alpha$ at $h=1$,  $\beta=1$.
The left panel shows the solutions for $\omega>0$ and the right one for $\omega<0$.}
    \label{fig_Omega_neq_0_sm}
\end{figure}

\begin{figure}[t!]
\begin{center}
    \includegraphics[width=1.\linewidth]{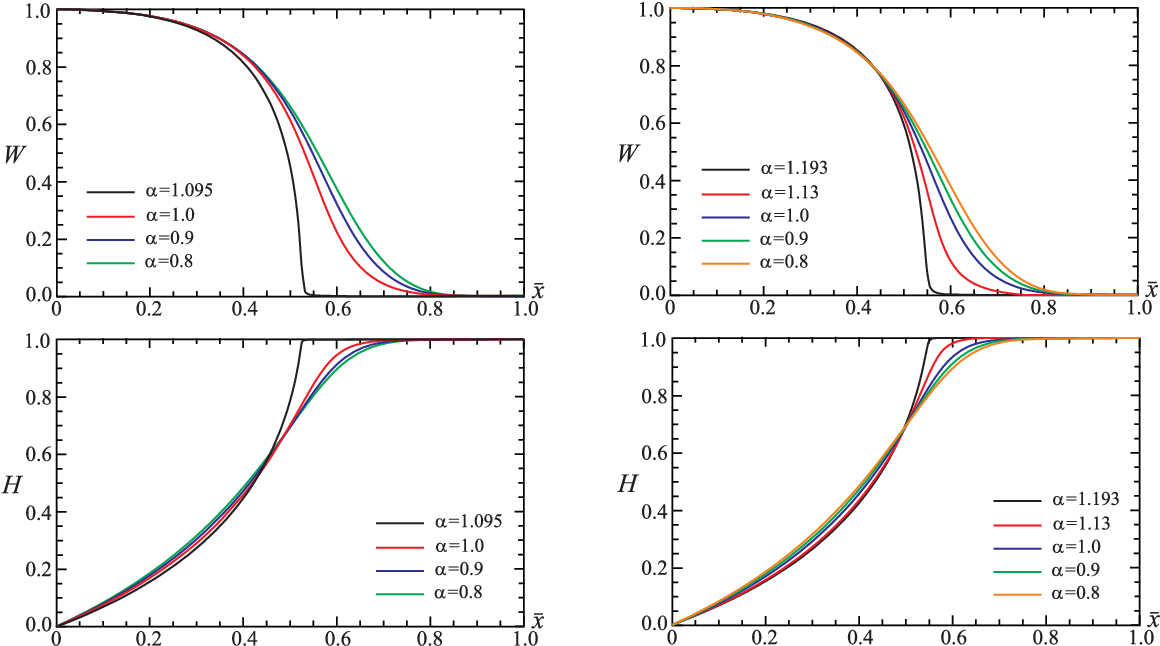}
\end{center}
\vspace{-0.5cm}
    \caption{The profiles of the gauge and scalar  functions of the solutions of the system~\re{field_eqs_curved_Einstein_1}-\re{psi_eqn_2}
are shown as functions of the compactified radial  coordinate $\bar x=x/(1+x)$ for some set of values of the
effective gravitational coupling $\alpha$ at $h=1$,  $\beta=1$.
The left panel shows the solutions for $\omega>0$ and the right one for $\omega<0$.}
    \label{fig_Omega_neq_0_gs}
\end{figure}

Apart from the zero mode, the system of equations~\re{field_eqs_curved_Einstein_1}-\re{psi_eqn_2} supports a tower of regular normalizable
solutions for fermionic modes with $\omega \neq 0, ~~|\omega| < |h/2|$. Here, both components $u$ and $v$
are non-zero, and for $h<0$ they posses at least one node while for $h>0$ they are nodeless.
These solutions can be obtained numerically, now we have to solve the full system of
coupled differential equations~\re{field_eqs_curved_Einstein_1}-\re{psi_eqn_2} imposing the boundary conditions \re{BC}. Note that this system is not
invariant with respect to inversion of the sign of $\omega$.
Indeed, it is seen in Figs.~\ref{fig_Omega_neq_0_sm} and~\ref{fig_Omega_neq_0_gs},
which display the metric components $N(x)\, ,\sigma(x)$ and the fields
$u(x)\,, v(x)\,, W(x)\,, H(x)$ for some set of
values of the gravitational coupling $\alpha $ and fixed $\beta=1$ and $h=1$, that,
as $\omega \to +0$ and $\omega \to -0$, the configurations approach the RN limit in a different way.

\begin{figure}[t]
\begin{center}
    \includegraphics[width=.6\linewidth]{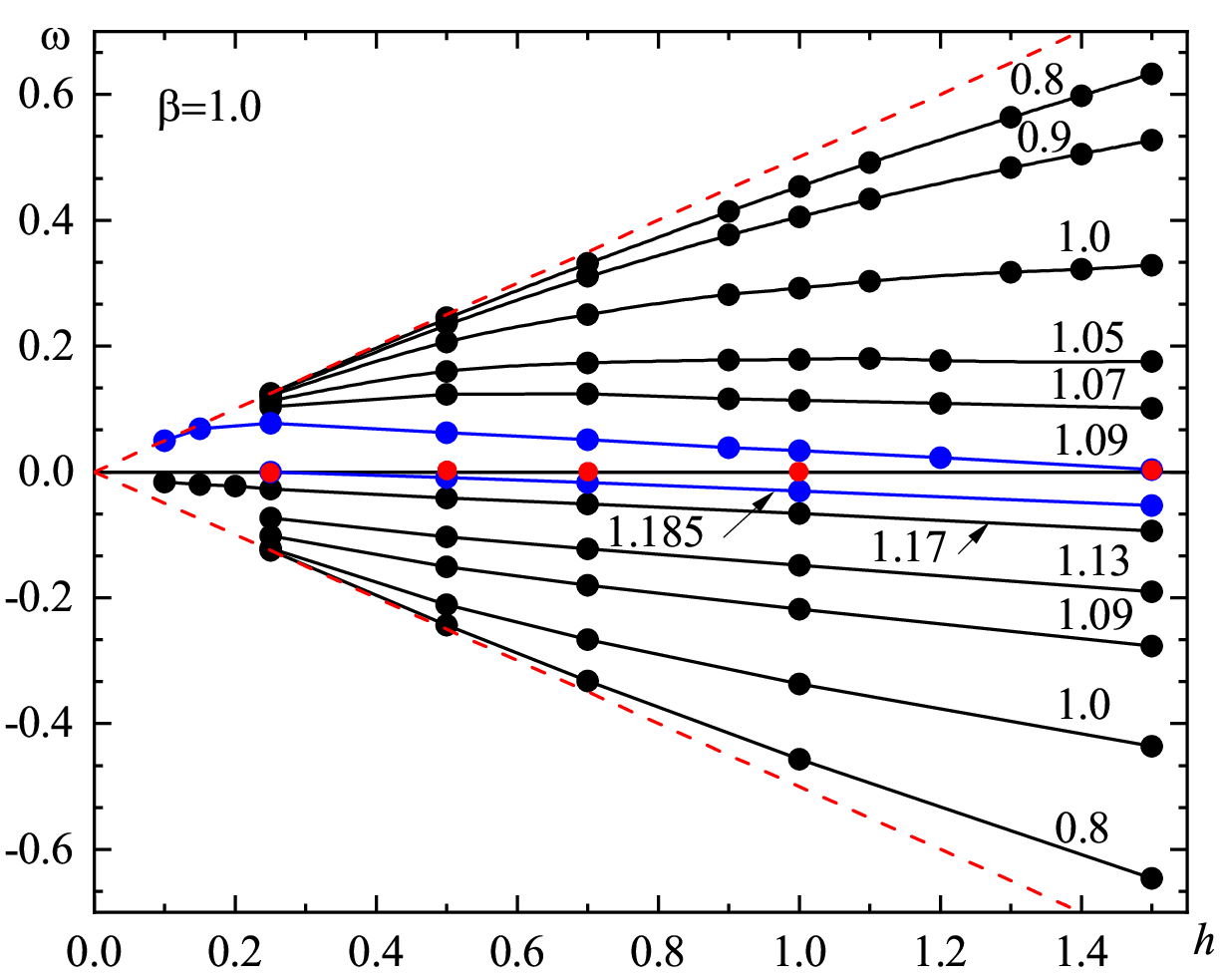}
\vspace{-0.5cm}
\end{center}
    \caption{
        Normalized energy of the localized fermionic states as a function of the Yukawa
        coupling $h$ for fixed $\beta=1$ and several values of $\alpha$ indicated by the numbers
        near the curves.
        The red dashed lines correspond to the continuum threshold $|\omega| = |h/2|$ in the limit
        $\alpha\to 0$. The blue lines correspond to the curves linked to the extremal RN black hole at some $h_{\text{cr}}$ as
        $\omega\to 0$. Bold red dots indicate the critical values $h_{\text{cr}}$ given in Table~\ref{tab_h_alpha}.}
    \label{fig_spectrum}
\end{figure}

In the flat space limit the fermion mode becomes delocalized as
$|\omega| \to |h/2|$, while increasing of the gravitational coupling stabilizes the system.
Both the ADM mass of the configuration and the eigenvalue $\omega$, which is defined from the numerical
calculations, are decreasing as $\alpha$ increases, see the right panel of Fig.~\ref{fig1}.
The evolution scenario depends generically on the values of the parameters of the model.
For example, setting  $\beta=1$ and $h=1$, we observe that there are two branches of solutions
which are linked to the negative and positive continuum: they
end at the  critical value $\alpha_{\text{cr}} \approx 1.095$ as $\omega \to +0$, and
at $\alpha_{\text{cr}}\approx 1.193$ as $\omega \to -0$
(cf. Fig.~\ref{fig_spectrum} and Table~\ref{tab_h_alpha}).
In both cases the
configuration reaches the embedded extremal RN solution \re{RN} in a way
which is qualitatively similar to that of the BPS monopole with localized
fermion zero mode discussed above.
As $\alpha$ tends to the critical
value, the eigenvalue $\omega$ approaches zero and the fermion field
is fully absorbed into interior of the forming black hole.

In Fig.~\ref{fig_spectrum} we plot the normalized energy of the localized fermionic
states as a function of the Yukawa coupling constant $h$. Having constructed some set of
solution for different values of $\alpha$, the following scenario becomes plausible.
As the Yukawa coupling increases from zero, while both $\beta$ and $\alpha$ are kept fixed,
a branch of normalizible non-zero fermion modes emerges smoothly from the self-gravitating monopole. The
energy of the localized fermionic states is restricted as $|\omega| < |h/2|$, as the
gravitational coupling remains relatively weak, the modes remain close to the
continuum threshold.

The spectral flow is more explicit as the coupling $\alpha$ becomes stronger, see Fig.~\ref{fig_spectrum}.
Increase of the Yukawa coupling, which yields the mass of the fermionic states, leads to increase of eigenvalues
$\omega$.  However, an interesting observation is that at some critical value of the parameter $h$, the energy of
the localized mode approaches some
maximal value. As the Yukawa coupling continue to grow, the corresponding eigenvalue
starts to decrease, it tends to zero as some maximal value $h_{\text{cr}}$. Again, in this limit the configuration
approaches the embedded RN solution \re{RN} and the fermion fields are again fully absorbed into interior of
the forming black hole.
The pattern is illustrated in Fig.~\ref{fig_spectrum}, where two blue curves
display the spectral flow of both positive and negative Dirac eigenvalues $\omega$ for  $\alpha=1.09$ and
$\alpha=1.185$, respectively. In the limiting case $\omega \to + 0$ one has $h_{\text{cr}}\approx 1.5$ (for $\alpha=1.09$)
and when $\omega \to - 0$ we have $h_{\text{cr}}\approx 0.25$ (for $\alpha=1.185$).

The general scenario is that, depending on the value of the Yukawa coupling constant $h$, there exist a
critical value of the gravitational coupling $\alpha_{\text{cr}}$ at which the spectral flow approaches the limit
$\omega \to \pm 0$ and the configuration runs to the embedded RN solution \re{RN}. Some corresponding values are given
in Table~\ref{tab_h_alpha}, and they are also displayed by the bold red dots in Fig.~\ref{fig_spectrum}.
Once again, each particular value of the Yukawa coupling gives rise to two distinct spectral flows approaching
the embedded RN solution as $\omega\to \pm 0$ at two different values of $\alpha_{\text{cr}}$.

\begin{figure}[t!]
\begin{center}
\includegraphics[width=1.\linewidth]{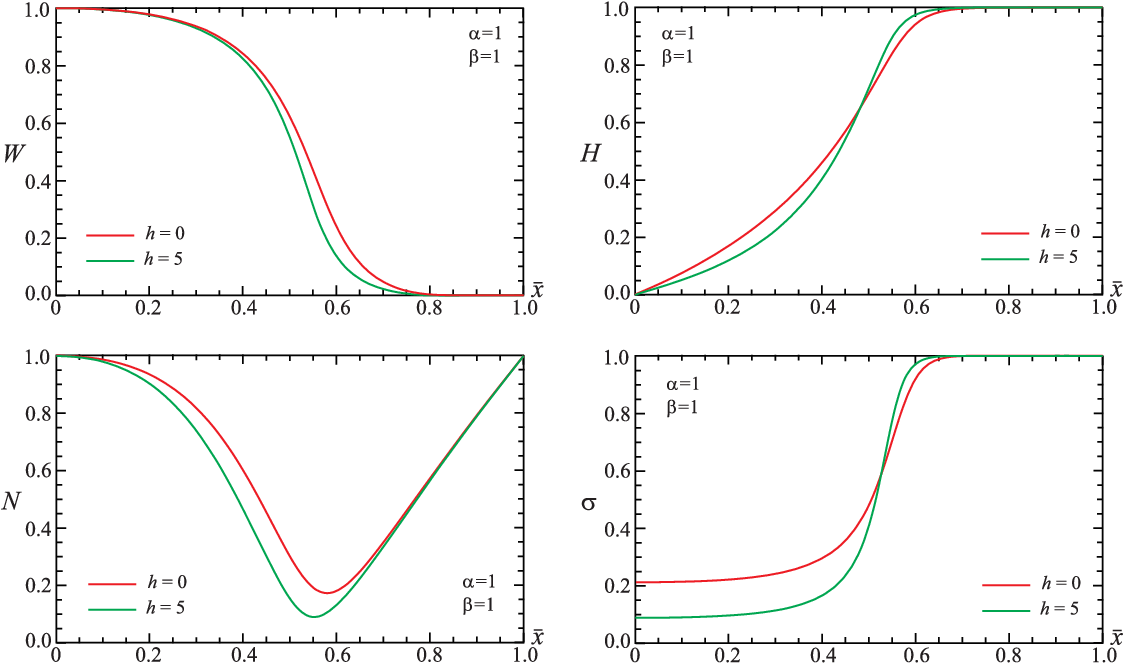}
\end{center}
\vspace{-0.5cm}
\caption{
The  profile functions of the gauge field $W(x)$, the scalar field $H(x)$, and the metric functions $N(x)$ and $\sigma(x)$
of the gravitating non-BPS monopole ($h=0$) and of the monopole-fermion system ($h= 5$)
are shown as functions of the compactified radial  coordinate $\bar x=x/(1+x)$ 
at $\alpha=1$ and $\beta=1$.
}
\label{fig6}
\end{figure}

\begin{table}[h!]
\scalebox{1.}{
\begin{tabular}{|c|c|c|c|c|c|}
    \hline
    \rule[-1ex]{0pt}{2.5ex}
    $h$&0.25&0.5&0.7&1.0&1.5\\
    \hline
    \rule[-1ex]{0pt}{2.5ex}
    $\alpha_{\text{cr}} (\omega\to +0)$&1.106&1.102&1.10&1.095&1.09\\
    \hline
    \rule[-1ex]{0pt}{2.5ex}
    $\alpha_{\text{cr}} (\omega\to -0)$&1.185&1.187&1.191&1.193&1.199\\
    \hline
\end{tabular}
}
\caption{
Critical values  $\alpha_{\text{cr}}$ at which $\omega\to \pm 0$ for some set of values of the Yukawa coupling
$h$ (cf. the red bold dots in Fig.~\ref{fig_spectrum}).
}
\label{tab_h_alpha}
\end{table}

Finally, we note that
the system of equations~\re{field_eqs_curved_Einstein_1}-\re{psi_eqn_2} possesses two characteristic limiting cases, $h\to \infty$ and $\beta\to \infty$.
First,  for
a fixed value of $\beta$ and increasing Yukawa coupling, the backreaction of the localized fermions becomes
stronger, the energy of the gravitating bounded fermionic mode increases and the profile functions of the monopole are
significantly deformed, see Fig.~\ref{fig6}. We observe that an increase of the
Yukawa coupling moves the configuration closer to the RN solution (see the bottom plots of Fig.~\ref{fig6}).
Note that deformations of the configuration
caused by its coupling with massive fermion modes may produce a number of interesting effects related with backreaction of
the fermions
\cite{Perapechka:2018yux,Perapechka:2019upv,Klimashonok:2019iya,Perapechka:2019vqv}.

Secondly, as the scalar field becomes very massive, the core of the monopole shrinks and in the limit
$\beta \to \infty$ the Higgs field is taking its vacuum expectation
value everywhere in space apart the origin.  One can expect that,
for the intermediate range of values of $\beta$, the scenario reported above for the $\beta=1$, should  persist.
Our numerical results confirm that an increase of the scalar coupling $\beta$ decreases the critical value of the
Yukawa coupling $h$ at which the configuration approaches the extremal Reissner-Nordstr\"{o}m solution.
However, for relatively large values of $\beta$,
the pattern of evolution of the self-gravitating monopole becomes different
\cite{Lee:1991vy,Breitenlohner:1991aa,Breitenlohner:1994di,Lue:1999zp,Kunz:2007jw}. One might expect also that the behavior of the fermion field
could be different in the large-$\beta$ regime.

\section{Conclusions}
The objective of this work is to investigate the fermionic modes
localized on the static spherically symmetric  self-gravitating non-Abelian monopole in the
$SU(2)$ Einstein-Dirac-Yang-Mills-Higgs theory.
We have constructed numerically solutions of the full system
of coupled field equations supplemented by the normalization condition
for the localized fermions, and investigated their properties. We have found that,
in addition to the usual zero mode, which always exists for a BPS monopole,
there is a tower of gravitationally localized states with
nonzero eigenvalues $\omega$, which are linked to the positive and
negative continuum.  While the fermionic zero mode exists for any negative value of the Yukawa coupling $h$, the
massive nodeless modes appear for positive values of $h$. We find that, as we increase the gravitational coupling,
the monopole bifurcates with the
extremal Reissner-Nordstr\"{o}m solution and the fermionic modes become absorbed into the interior of the forming black hole.
This scenario is viable for both zero and non-zero fermionic modes. Further, we observe that the Yukawa interaction
breaks the symmetry between the localized massive modes with positive and negative
eigenvalues. Another observation is that the
localized gravitating fermions may deform the monopole affecting the transition to the limiting solution.

The work here should be taken further by considering higher massive localized
fermionic states with some number of radial nodes. Another interesting
question, which we hope to be addressing in the near
future, is to investigate the effect of the bare mass of the fermions, localized on the  monopole.
Another direction can be related with investigation of properties
of charged fermions localized on the self-gravitating dyon.
Finally, let us note that there can be several fermionic modes localized by the gravitating monopole.
We hope to address these problems in our future work.

\section*{Acknowledgments}
Y.S. would like to thank Jutta Kunz  and Michael Volkov  for enlightening discussions. He
gratefully acknowledges the support of the Alexander von Humboldt Foundation and HWK Delmenhorst.
The work was supported by the Science Committee of the Ministry of Science and Higher Education of the Republic of Kazakhstan
(Grant  No. AP14869140, ``The study of QCD effects in non-QCD theories'')

\appendix

\section{Definition of the $\gamma^5$ matrix in curved (3+1)-dimensional spacetime}
\label{append1}

The interaction term between the spin-isospin fermions and the Higgs field of
non-Abelian self-gravitating monopole in the Lagrangian \re{Lag-spinor} is
\begin{equation}
    - \imath h \bar \psi^i_\alpha \left( \tilde{\gamma}^5 \right)_{\alpha \beta} \sigma^a_{i j }
    \phi^a \psi^j_{\beta} \, ,
\label{D_10}
\end{equation}
where  $\tilde{\gamma}^5$ is defined in the curved (3+1)-dimensional spacetime as
\begin{equation}
    \tilde{\gamma}^5 = \frac{1}{4!} E_{\alpha \beta \rho \sigma}
    \gamma^\alpha \gamma^\beta \gamma^\rho \gamma^\sigma =
    \frac{1}{4!} \sqrt{- g} \epsilon_{\alpha \beta \rho \sigma} e_a^\alpha e_b^\beta e_c^\rho e_d^\sigma
    \gamma^a \gamma^b \gamma^c \gamma^d =
    \frac{1}{4!} \sqrt{- g} \left(
        \epsilon_{\alpha \beta \rho \sigma} \epsilon^{a b c d} e_a^\alpha e_b^\beta e_c^\rho e_d^\sigma
    \right) \gamma^5 \, .
\label{D_20}
\end{equation}
Here
$
    E_{\alpha \beta \rho \sigma} = \sqrt{- g} \epsilon_{\alpha \beta \rho \sigma}
$ is the Levi-Civita tensor in curved space,
$
    \epsilon_{\alpha \beta \rho \sigma}
$ is the Levi-Civita tensor in flat space, and
$$
    \gamma^a \gamma^b \gamma^c \gamma^d = \epsilon^{a b c d} \gamma^5 , \quad
    \gamma^5 = \imath \gamma^0 \gamma^1 \gamma^2 \gamma^3 .
$$
The expression in the round brackets in \eqref{D_20} is the determinant of the matrix  $e^\alpha_a$:
$$
    \frac{1}{4!}
    \epsilon_{\alpha \beta \rho \sigma} \epsilon^{a b c d} e_a^\alpha e_b^\beta e_c^\rho e_d^\sigma =
    \det\left( e^\alpha_a \right) = \frac{1}{\sqrt{-g}}.
$$
Hence, the interaction term~\eqref{D_10} can be written as
$$
    - \imath h \bar \psi^i_\alpha \left( \gamma^5 \right)_{\alpha \beta} \sigma^a_{i j } \phi^a \psi^j_{\beta} .
$$

\end{document}